\documentclass[final,5p,times,twocolumn]{elsarticle}
\pdfoutput=1
\usepackage{amsmath}
\usepackage{graphicx}
\usepackage{dcolumn}

\usepackage{amssymb}
\usepackage{bm}

\newcommand{\be}{\begin{equation}}
\newcommand{\ee}{\end{equation}}
\newcommand{\bea}{\begin{eqnarray}}
\newcommand{\eea}{\end{eqnarray}}

\begin{document}

\begin{frontmatter}


  
\title{Multi-level Monte Carlo computation of the hadronic vacuum\\
       polarization contribution to $(g_\mu-2)$}

\author[a,b]{Mattia Dalla Brida} 
\author[a,b]{Leonardo Giusti}
\author[a,b]{Tim Harris}
\author[b]{Michele Pepe}

\address[a]{Dipartimento di Fisica, Universit\`a di Milano-Bicocca\\ Piazza della Scienza 3, 
  I-20126 Milano, Italy}

\address[b]{INFN, Sezione di Milano-Bicocca \\ Piazza della Scienza 3, 
I-20126 Milano, Italy}

\begin{abstract}
The hadronic contribution to the muon anomalous magnetic moment $a_\mu=(g_\mu-2)/2$
has to be determined at the per-mille level for the Standard Model prediction
to match the expected final uncertainty from the ongoing E989 experiment.
This is 3 times better than the current precision from the dispersive approach, and 5-15
times smaller than the uncertainty on the purely theoretical determinations from lattice QCD.
So far the stumbling-block is the large statistical error in the Monte Carlo evaluation of the
required correlation functions which can hardly be tamed by brute force. Here we propose to solve
this problem by multi-level Monte Carlo integration, a technique which reduces the variance of
correlators exponentially in the distance of the fields. We test our strategy by computing the 
Hadronic Vacuum Polarization on a lattice with a linear extension of 3~fm, a spacing of 0.065 fm,
and a pion mass of 270 MeV. Indeed the two-level integration makes the contribution to the statistical
error from long-distances de-facto negligible by accelerating its inverse scaling with the cost
of the simulation. These findings establish multi-level Monte Carlo as a solid and efficient
method for a precise lattice determination of the hadronic contribution to $a_\mu$. As the approach is
applicable to other computations affected by a signal-to-noise ratio problem, it has the potential to
unlock many open problems for the nuclear and particle physics community.
\end{abstract}






\end{frontmatter}


\section{Introduction}
\vspace{-0.25cm}

The current experimental value of the muon anomalous magnetic moment $a_\mu = 116 592 08.9(6.3)\times 10^{-10}$
by the E821 experiment has the remarkable precision of $0.54$ parts per million (ppm) \cite{Bennett:2006fi}, while
the on-going E989 experiment at FNAL is expected to reach the astonishing precision of $0.14$ ppm by the
end of its operation~\cite{Grange:2015fou} when also E34 at J-PARC may be well under way~\cite{Abe:2019thb}.
The Standard Model (SM)
prediction includes contributions
from five-loop Quantum Electrodynamics, two-loop Weak interactions, the Hadronic leading-order
Vacuum Polarization (HVP) and
(the much smaller) Hadronic Light-by-Light scattering (HLbL), see
Ref.~\cite{Aoyama:2020ynm} and references therein.
The overall theoretical uncertainty is dominated by the hadronic part.
So far\footnote{A recent proposal for an independent determination of the HVP
from the muon-electron elastic scattering has been put forward in Ref.~\cite{Abbiendi:2016xup}.}, 
lacking precise purely theoretical computations, the hadronic contributions have been extracted (by assuming the SM)
from experimental data via dispersive integrals (HVP \& HLbL) and low-energy effective models supplemented with the
operator product expansion (HLbL). This leads to $a_\mu = 116 591 81.0(4.3)\times 10^{-10}$
(0.37 ppm)~\cite{Aoyama:2020ynm}, which deviates by $3-4$ standard deviations from the E821 result,
a difference persisting for a decade which may be a hint for a New Physics signal.

State-of-the-art lattice Quantum Chromodynamics (QCD) determinations of the HVP are becoming competitive.
At present, quoted uncertainties range between
$0.6\%$ to roughly $2\%$ \cite{Blum:2018mom,Giusti:2019xct,Davies:2019efs,Shintani:2019wai,Gerardin:2019rua,Aubin:2019usy,Borsanyi:2020mff}
corresponding to an overall error on $a_\mu$ which is still 5-15 times larger than the anticipated uncertainty
from E989. Taken at face value, the most recent lattice determination of the HVP \cite{Borsanyi:2020mff} differs
from the dispersive result by more than 3 standard deviations, and generates tensions with the
global electroweak fits~\cite{Passera:2008jk,Keshavarzi:2020bfy,Crivellin:2020zul}.  

All these facts call for an independent theoretically-sound lattice computation of the hadronic
contribution to $a_\mu$ at the per-mille level from first principles. The main bottleneck toward this goal 
is the large statistical error in the Monte Carlo evaluation of the required correlation functions, see
Ref.~\cite{Aoyama:2020ynm} and references therein.
The aim of this letter is to solve
this problem by a novel
computational paradigm based on multi-level Monte Carlo
integration in the presence of fermions~\cite{Ce:2016idq,Ce:2016ajy,DallaBrida:2020prep}. With respect to the
standard approach, this strategy reduces the variance exponentially with the temporal distance of the fields,
thus opening the
possibility of making negligible the contribution
to the statistical error from long-distances. Here we focus on the HVP, but the strategy is general and can
be applied to the HLbL, the isospin-breaking and electromagnetic contributions as well.
\vspace{-0.25cm}

\section{The signal-to-noise problem}
The HVP can be written as~\cite{Bernecker:2011gh}
\be\label{eq:amuint}
a_\mu^{\rm HVP} = \left(\frac{\alpha}{\pi}\right)^2 \int_0^{\infty} d x_0\, K(x_0,m_\mu) \, G(x_0)\; ,
\ee
where $\alpha$ is the electromagnetic coupling constant, $K(x_0,m_\mu)$ is a known analytic
function which increases quadratically at large $x_0$~\cite{DellaMorte:2017dyu}, $m_\mu$ is the muon mass,
and $G(x_0)$ is the zero-momentum correlation function 
\be\label{eq:correlat}
 G(x_0) = \int d^3 {\bf x}\, \langle J_k^{em}(x) J_k^{em}(0) \rangle
\ee
of two electromagnetic currents $J_k^{em}= i \sum_{i=1}^{N_f} q_i \bar\psi_i\gamma_k\psi_i$. In this
study we consider $N_f=3$, the three lighter quarks of QCD with the first two
degenerate in mass, so that
\be
G(x_0) = G^{\rm conn}_{u,d}(x_0) + G^{\rm conn}_{s}(x_0) +  G^{\rm disc}_{u,d,s}(x_0)\; .  
\ee
The light-connected Wick contraction $G^{\rm conn}_{u,d}(x_0)$ and the disconnected
one $G^{\rm disc}_{u,d,s}(x_0)$ are the most problematic contributions with regard to the
statistical error. In standard Monte Carlo computations, the relative error of the former
at large time distances $|x_0|$ goes as  
\be\label{eq:relerr}
\frac{\sigma^2_{_{G^{\rm conn}_{\rm u,d}}}(x_0)}{[G^{{\rm conn}}_{\rm u,d}(x_0)]^2}\propto
\frac{1}{n_0}\; e^{2\, (M_\rho - M_\pi)|x_0|}\; ,  
\ee
where $M_\rho$ is the lightest asymptotic state in the iso-triplet vector
channel,
and $n_0$ is the number of independent field configurations. For physical
values of the quark masses, the difference $(M_\rho-M_\pi)$
can be as large as $3.2$~fm$^{-1}$. The computational effort, proportional to $n_0$,
of reaching a given relative statistical error thus increases exponentially
with the distance $|x_0|$. For the disconnected
contribution $G^{\rm disc}_{u,d,s}(x_0)$, the situation is even worse since the
variance is constant in time and therefore the coefficient multiplying
$|x_0|$ is larger.
At present this exponential increase of the relative error is the barrier
which prevents lattice theorists to reach a per-mille statistical precision
for the HVP. In order to mitigate this problem, in state-of-the-art calculations the
contribution to the integral in Eq.~(\ref{eq:amuint}) is often computed from Monte
Carlo data only up to time-distances of $1.5-2$~fm or so, while the rest is
estimated by modeling $G(x_0)$, see Ref.~\cite{Gulpers:2020pnz} for an
up-to-date review.

\section{Multi-level Monte Carlo}
\begin{figure}[!t]
\includegraphics[width=8.5 cm,angle=0]{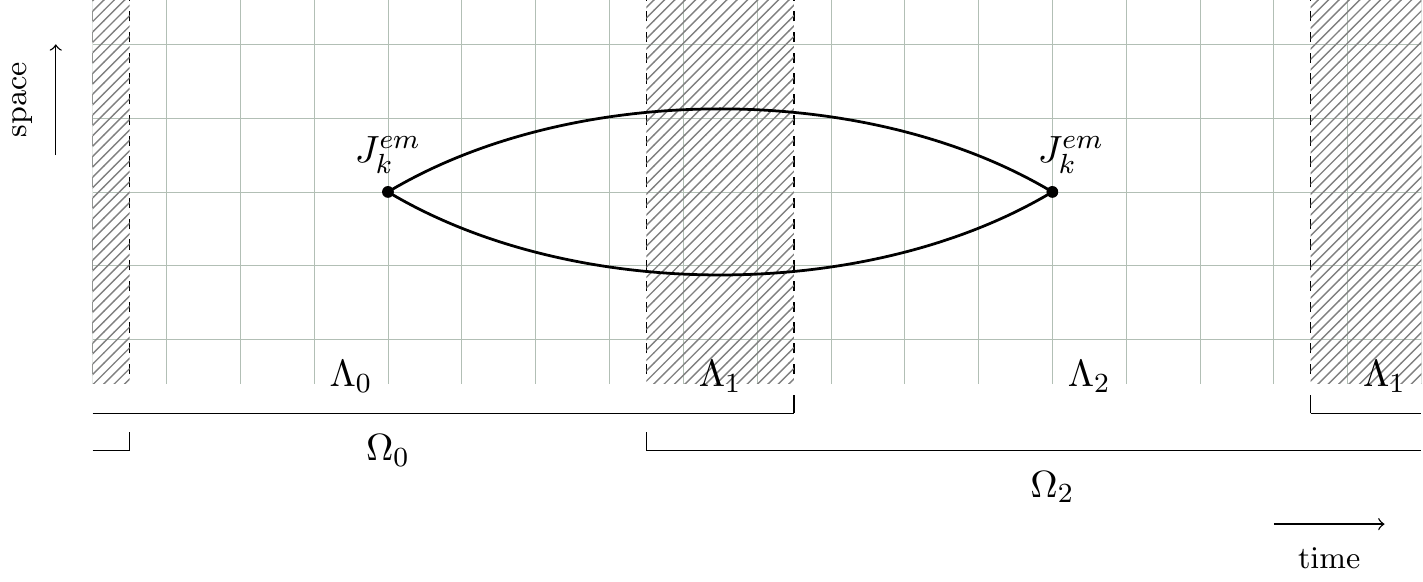}
\caption{\label{fig:MB-DD} Domain decomposition of the lattice adopted in this paper.
Periodic (anti-periodic) boundary conditions in the time direction are enforced for
gluons (fermions).}
\end{figure}

Thanks to the conceptual, algorithmic and technical progress over the last few years,
it is now possible to carry out multi-level Monte Carlo simulations in the presence of
fermions~\cite{Ce:2016idq,Ce:2016ajy}. The first step in this approach is the decomposition of the
lattice in two overlapping domains $\Omega_0$ and $\Omega_2$, see e.g. Fig.~\ref{fig:MB-DD},
which share a common region $\Lambda_1$. The latter is chosen so that the minimum distance between the
points belonging to the inner domains $\Lambda_0$ and $\Lambda_2$ remains finite
and positive in the continuum limit.\\
\indent The next step consists in rewriting the determinant of the Hermitean massive Wilson-Dirac
operator $Q=\gamma_5 D$ as
\begin{equation}\label{eq:factfinal0}
  \det\, Q= \frac{\det\, \left(1-w\right)}{\det\, Q_{\Lambda_{1}}
    \det\, Q^{-1}_{\Omega_0} \det\, Q^{-1}_{\Omega_2}} \; ,  
\end{equation}
where $Q_{\Lambda_{1}}$, $Q_{\Omega_0}$, and $Q_{\Omega_2}$ indicate the very same 
operator restricted to the domains specified by the subscript. They are obtained from $Q$
by imposing Dirichlet boundary conditions on the external boundaries of each domain.
The matrix $w$ is 
\begin{equation}\label{eq:w}
w=P_{\partial\Lambda_{0}} Q^{-1}_{\Omega_0}\, Q_{\Lambda_{1,2}}\, Q^{-1}_{\Omega_2} Q_{\Lambda_{1,0}} \, ,
\end{equation}  
where $Q_{\Lambda_{1,0}}$ and $Q_{\Lambda_{1,2}}$ are the hopping terms of the operator $Q$ across the
boundaries in between the inner domains $\Lambda_0$ and $\Lambda_2$ and the common region
$\Lambda_1$ respectively, while $P_{\partial\Lambda_{0}}$ is the projector on the inner boundary of
$\Lambda_0$~\cite{Ce:2016ajy}. The denominator in Eq.~(\ref{eq:factfinal0})
has already a factorized
dependence on the gauge field since $\det Q_{\Lambda_{1}}$, $\det\, Q^{-1}_{\Omega_0}$
and $\det\, Q^{-1}_{\Omega_2}$ depend only on the gauge field in $\Lambda_1$, 
$\Omega_0$ and $\Omega_2$ respectively.
\begin{figure*}[thb]
\vspace{-1.5cm}

\begin{center}
\begin{tabular}{cc}
\includegraphics[width=8.0 cm,angle=0]{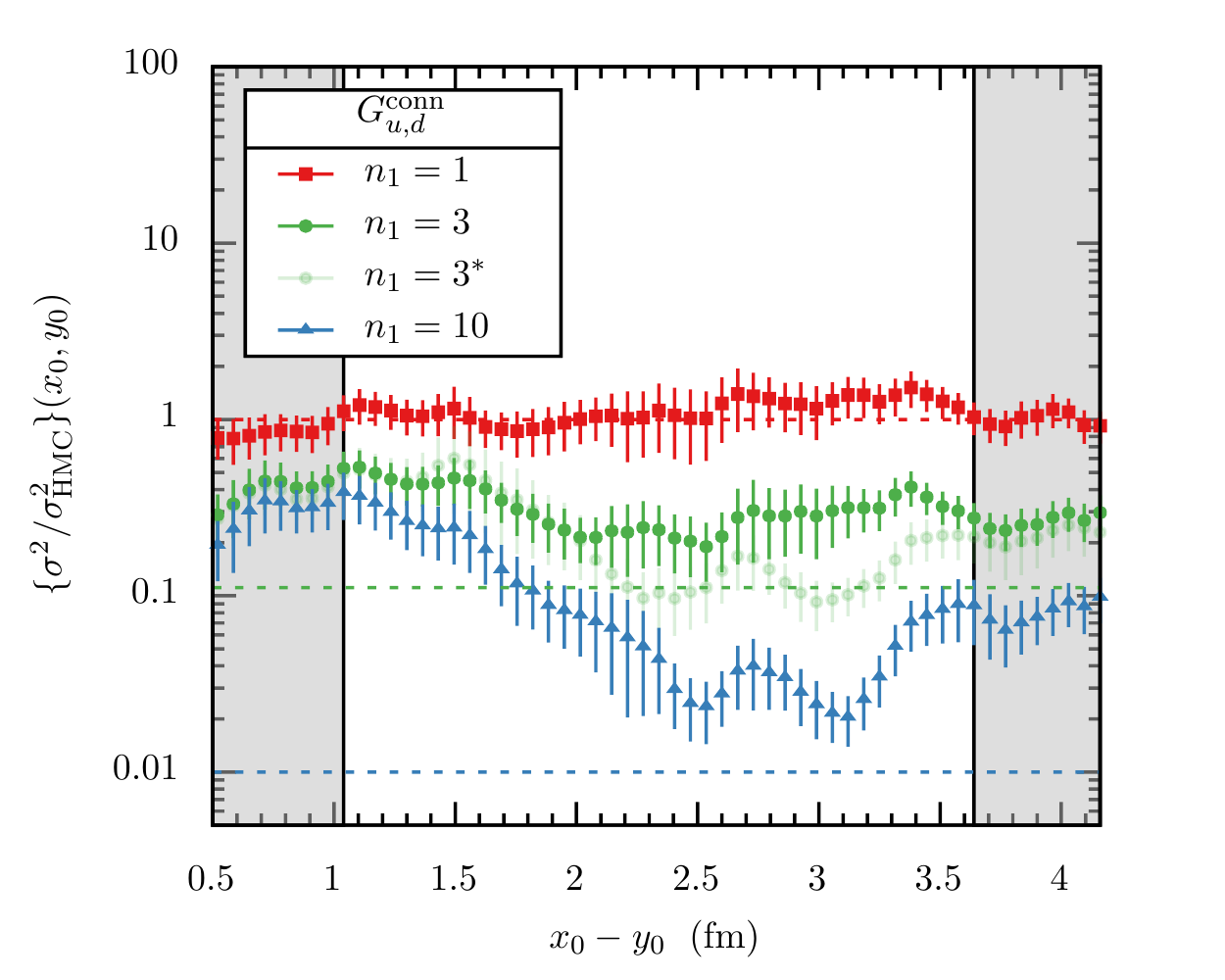} &
\includegraphics[width=8.0 cm,angle=0]{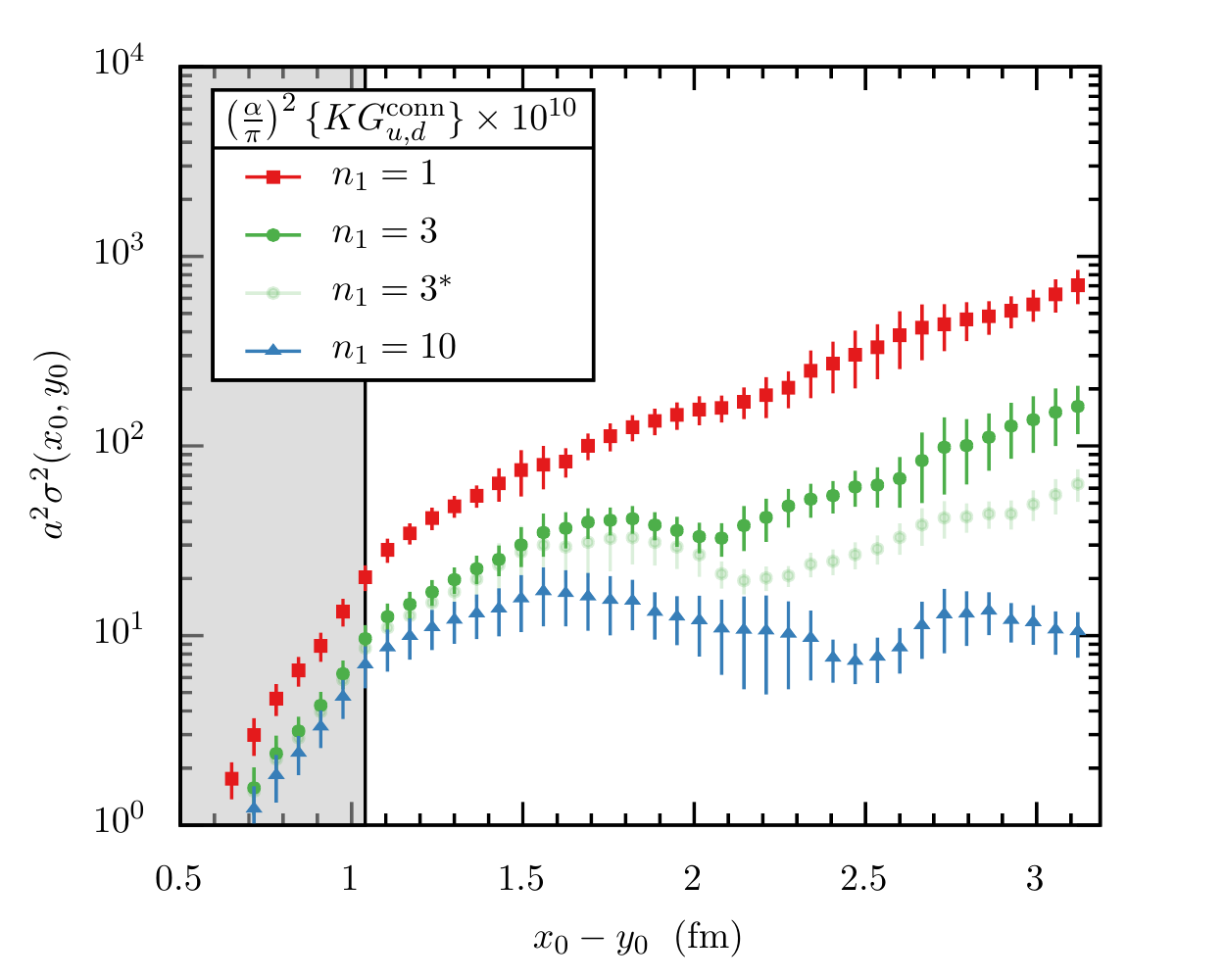}\\
\end{tabular}
\vspace{-0.5cm}

\caption{Left: variance of the light-connected contraction as a
function of the difference between the time-coordinates of the currents for
$n_1=1,3,10$ ($n_1=3^*$ is obtained by skipping $48$~MDUs between two consecutive
level-$1$ configurations). Data are normalized to the analogous ones computed on CLS
configurations generated by one-level HMC. Dashed lines represent the maximum reduction
which can be obtained by two-level integration, namely $1/n^2_1$, in the absence of
correlations between level-$1$ configurations. Grey bands indicate the thick time-slices
where the gauge field is kept fixed during level-$1$ updates. Right: variance of the light-connected
contribution to the integrand in Eq.~(\ref{eq:amuint}).
\label{fig:variances}}
\end{center}
\vspace{-0.75cm}

\end{figure*}

In the last step, the numerator in Eq.~(\ref{eq:factfinal0}) is rewritten as
\be\label{eq:MB}
\det\, \left(1-w\right) =
\frac{\det\,[1-R_{N+1}(1-w)]}{C \prod_{k=1}^{N/2} {\det} \big[ (u_k -w )^\dagger (u_k -w) \big]}\, ,
\ee
where $u_k$ and $u^*_k$ are the $N$ roots of a polynomial approximant for $(1-w)^{-1}$,
the numerator is the remainder, and $C$ is an irrelevant constant. The denominator
in Eq.~(\ref{eq:MB}) can be represented
by an integral over a set of $N/2$ multi-boson fields~\cite{Luscher:1993xx,Borici:1995np}
having an action with a factorized dependence on the gauge field in $\Lambda_0$ and
$\Lambda_2$ ~\cite{Ce:2016idq,Ce:2016ajy,DallaBrida:2020prep} inherited from $w$.
When the polynomial approximation is properly chosen, see below, the
remainder in the numerator of Eq.~(\ref{eq:MB}) has mild fluctuations in the gauge field, and is
included in the observable in the form of a reweighting factor in order to obtain
unbiased estimates.

A simple implementation of these ideas is to divide the lattice as shown
in Fig.~\ref{fig:MB-DD}, where $\Lambda_0$ and $\Lambda_2$ have the shape of thick
time-slices while $\Lambda_1$ includes the remaining parts of the lattice.
The short-distance suppression of the quark propagator implies that
a thickness of $0.5$~fm or so for the thick-time slices forming $\Lambda_1$
is good enough, see e.g. Fig.~4 in Ref.~\cite{Luscher:2003vf}, and is not expected
to vary significantly with the quark mass. This is the domain decomposition that
we use for the numerical computations  presented in this letter.

The Monte Carlo simulation is then performed using a two-level scheme.
We first generate $n_0$ level-$0$ gauge field configurations by updating the field over the entire lattice;
then, starting from each level-$0$ configuration, we keep fixed the gauge field in the overlapping
region $\Lambda_1$, and generate $n_1$ level-$1$ configurations by updating 
the field in $\Lambda_0$ and in $\Lambda_2$ independently thanks to the factorization of the action.
The resulting gauge fields are then combined
to obtain effectively $n_0\cdot n_1^2$ configurations at the cost of generating $n_0\cdot n_1$
gauge fields over the entire lattice. In particular, for each level-$0$ configuration,
we compute the statistical estimators
by averaging the values of the correlators over the $n_1^2$ level-$1$ gauge fields.
Previous experience on two-level integration~\cite{Luscher:2001up,DellaMorte:2010yp,Ce:2016idq,Ce:2016ajy}
suggests that, with two independently updated regions, the variance decreases proportionally
to $1/n_1^2$ until the standard deviation of the
estimator is comparable with the signal, i.e. until the level-$1$ integration has solved the signal-to-noise problem.
From Eq.~(\ref{eq:relerr}) we thus infer that the variance reduction due to level-$1$ integration
is expected to grow exponentially with the time-distance of the currents in Eq.~(\ref{eq:correlat}).
The overhead for simulating the extra multi-boson fields increases the cost by an overall
constant factor which is quickly amortized by the improved scaling.

\begin{figure*}[thb]
\vspace{-1.5cm}

\begin{center}
\begin{tabular}{cc}
\includegraphics[width=8.0 cm,angle=0]{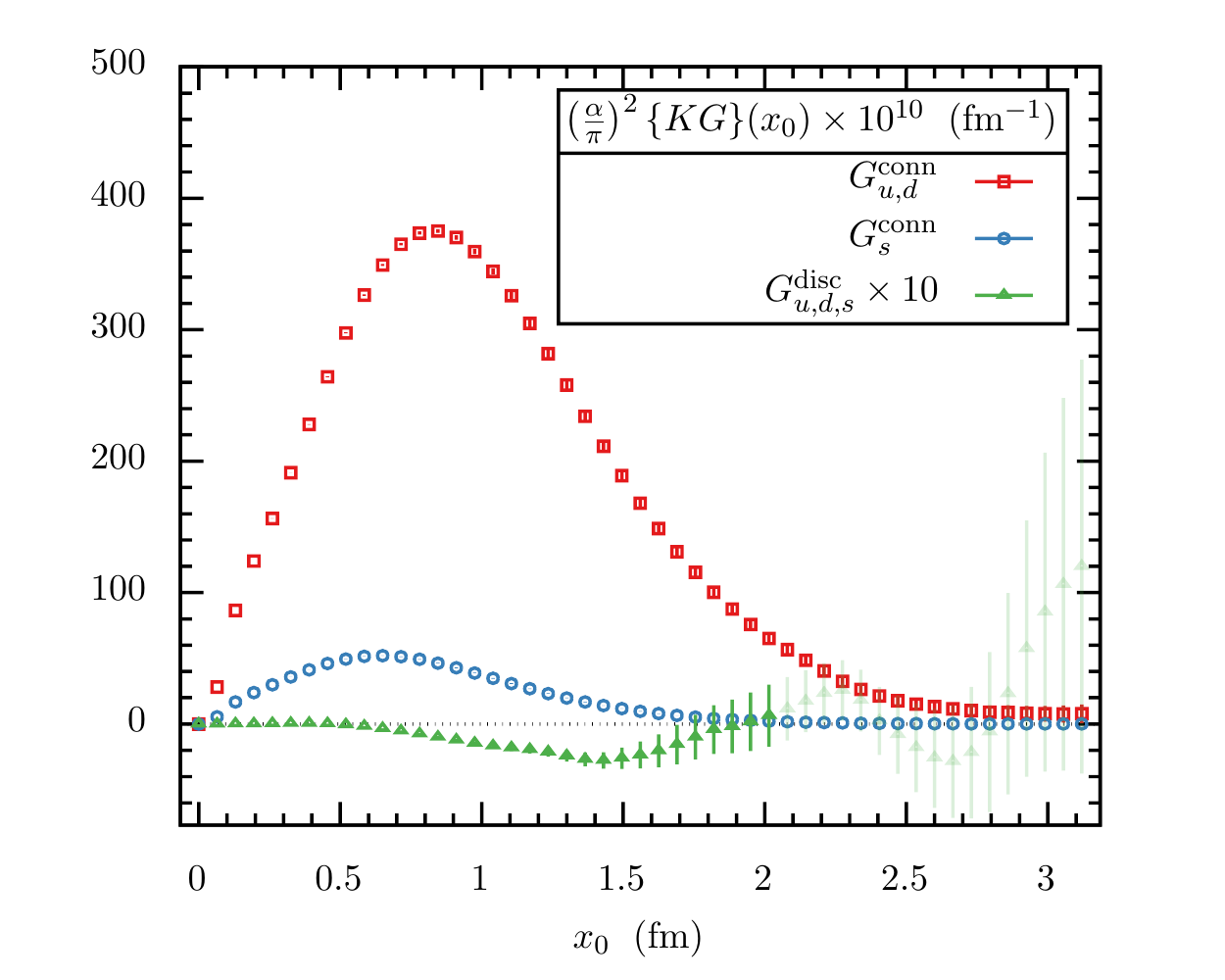} &
\includegraphics[width=8.0 cm,angle=0]{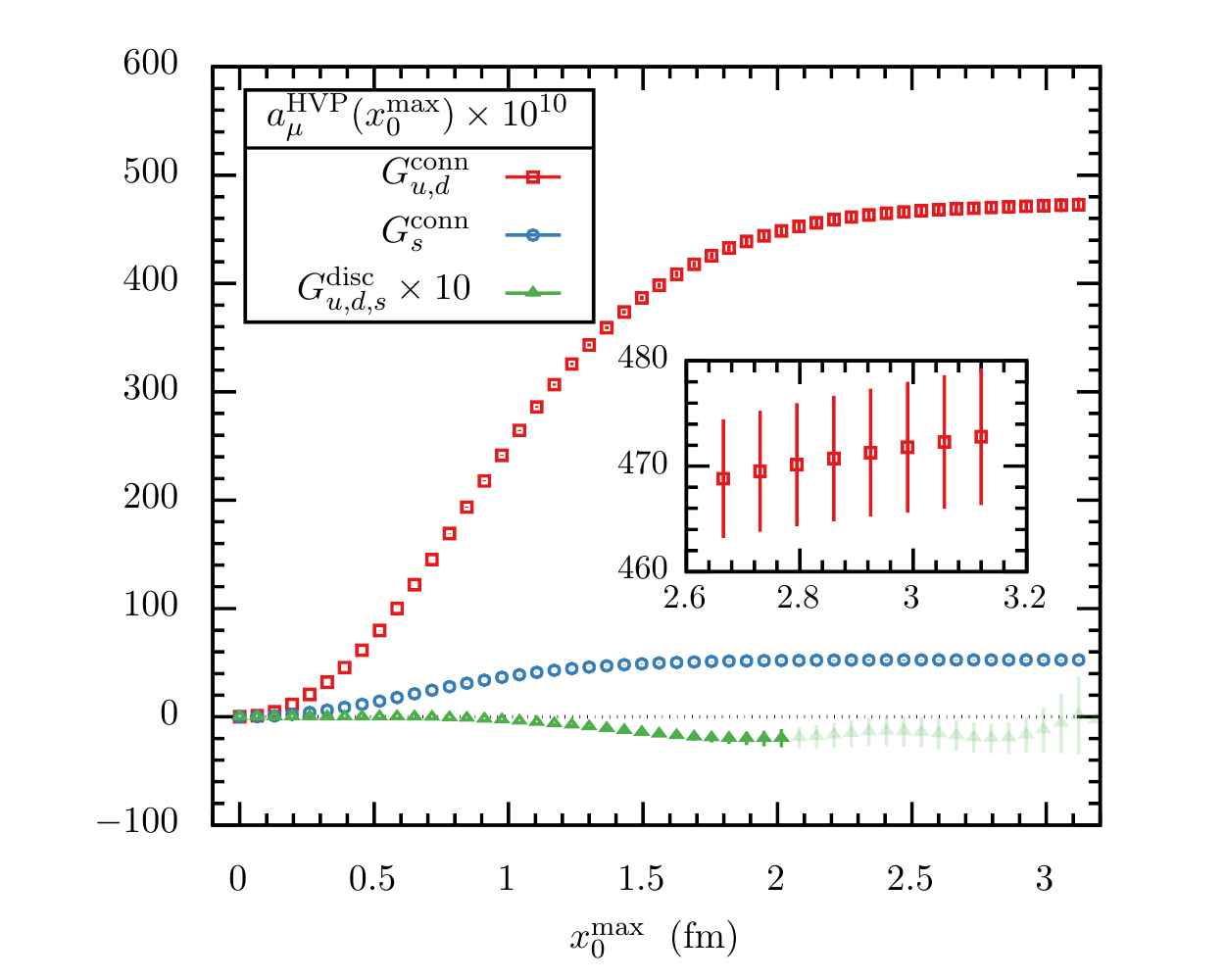}\\
\end{tabular}
\vspace{-0.5cm}

\caption{Left: best results for the contribution to the integrand in Eq.~(\ref{eq:amuint})
from the light-connected (red squares), strange-connected (blue circles) and disconnected (green triangles) contractions as
a function of the time coordinate. Right: best results for the contributions to
$a_\mu^{\rm HVP}$ from light-connected (red squares), strange-connected (blue circles), and disconnected (green triangles)
contractions as a function of the upper extrema of integration $x_0^{\rm max}$.
\label{fig:integrand}}
\end{center}
\vspace{-0.75cm}

\end{figure*}

\section{Lattice computation}
In order to assess the potential of two-level Monte Carlo integration, we simulate QCD with two
dynamical flavours supplemented by a valence strange quark. Gluons are discretized
by the Wilson action while quarks by the $O(a)$-improved Wilson--Dirac operator, see
Refs.~\cite{Engel:2014eea,DallaBrida:2020prep} for unexplained definitions.
Periodic and anti-periodic boundary conditions are imposed on the gluon and fermion fields
in the time direction respectively, while periodic conditions are chosen
for all fields in the spatial directions. We simulate a lattice of size $96\times 48^3$ with an
inverse bare coupling constant $\beta=6/g_0^2=5.3$, corresponding to a spacing of
$\;a=0.065$\,fm \cite{Fritzsch:2012wq,Engel:2014eea}. The size of the lattice, rather large for
a proof of concept, is chosen so to be able to accommodate a light pion and still be in the
large volume regime, namely $M_\pi=270$~MeV and $M_\pi L\geq4$. The domains $\Lambda_0$ and
$\Lambda_2$ are the union of $40$ consecutive time-slices, while each thick time-slice forming
the overlapping region $\Lambda_1$ is made of $8$ time-slices corresponding to a thickness
of approximately $0.5$~fm. The determinants in the
denominator of Eq.~(\ref{eq:factfinal0}) are taken into account by standard pseudofermion
representations, while the number of multi-bosons is fixed to $N=12$. The very same action and
set of auxiliary fields are used either at level-$0$ or level-$1$. The reweighting
factor is estimated stochastically with 2 random sources, which are enough for its contribution
to the statistical error to be negligible. Further details on the algorithm
and its implementation can be found in Ref.~\cite{DallaBrida:2020prep}.

We generate $n_0=25$ level-0 configurations separated by $48$ molecular dynamics units (MDU),
so that in practice they can be considered statistically
uncorrelated~\cite{Fritzsch:2012wq,Engel:2014eea}. For each level-$0$ background gauge field, 
we generate $n_1=10$ configurations in $\Lambda_0$ and $\Lambda_2$ spaced by $16$~MDUs.
The connected contraction is calculated by inverting the Wilson-Dirac operator
on local sources, while the disconnected one is computed via split-even random-noise
estimators~\cite{Giusti:2019kff}. For each level-$0$ configuration, the statistical
estimators are computed by averaging the correlators over the $n_1^2$ combinations of level-$1$ fields.
The error analysis then proceeds as usual. For the sake of the presentation, we show results in
physical units and properly renormalized: the central value of the lattice spacing is taken from
Ref.~\cite{Engel:2014eea}, and the one of the vector-current renormalization constant
from Ref.~\cite{DallaBrida:2018tpn}. We do not take into account their contributions to
the errors since on one side we are interested in investigating
the statistical precision of the vector correlator computed via two-level integration only,
and on the other side the numerical accuracy of those quantities can be improved independently.

To single out the reduction of the variance due only to two-level averaging, we carry out a
dedicated calculation of correlation functions. We compute the
light-connected contraction by averaging over $216$ local sources put on the time-slice ($y_0/a=32$)
of $\Lambda_0$ at a distance of $8$ lattice spacings from its right boundary and, as usual,
by summing over the sink space-position. This large number of sources
guarantees that the dependence of the variance on the gauge field in the domains $\Lambda_0$
and $\Lambda_2$ is on equal footing, since no further significant variance reduction
is observed by increasing their number. We determine the disconnected contraction by
averaging each single-propagator trace over a large number of Gaussian random sources,
namely $768$, so to have a negligible random-noise contribution to the variance~\cite{Giusti:2019kff}.

The variance of the light-connected contribution as a function of the distance from the source is shown
on the left plot of Fig.~\ref{fig:variances}. For better readability only the time-slices belonging
to $\Omega_2$ are shown, i.e. those relevant for studying the effect of two-level integration
given the source position. Data are normalized to the variance obtained with the same number of sources
on CLS configurations\footnote{https://wiki-zeuthen.desy.de/CLS/CLS.} which were generated with a 
conventional one-level HMC~\cite{DelDebbio:2006cn,DelDebbio:2007pz,Fritzsch:2012wq}.
The exponential reduction of the variance with the distance from the source is manifest in the data,
with the maximum gain reached from $2.5$~fm onward for $n_1=10$. The loss of about a factor between $2$ and
$3$ with respect to the best possible scaling, namely $n_1^2$, either for $n_1=3$ or $10$ (dashed lines) is
compatible with the presence of a residual correlation among level-$1$ configurations. Indeed the
variance reduction for $n_1=3$, obtained by skipping $48$~MDUs between consecutive level-$1$
configurations (labeled by $n_1=3^*$), is compatible with the $n_1^2$ scaling at
large distances within errors. In our particular setup, even for $n_1=10$ the statistical error at large
distances scales de-facto with the inverse of the cost of the simulation rather than with its squared
root. This is easily seen by comparing the variance reduction shown in the left plot
of Fig.~\ref{fig:variances} with the cost of the simulation for $n_1=10$. The latter is   
in fact $4$ times the one for $n_1=1$ due to the different separation in MDU units
between two consecutive configurations at level-$0$ and level-$1$.\\
\indent The power of the two-level integration can be better appreciated from the right plot
of Fig.~\ref{fig:variances},
where we show the variance of the light-connected contribution to the integrand in Eq.~(\ref{eq:amuint})
as a function of the time-distance of the currents. {The sharp rising of the variance}
computed by one-level Monte Carlo ($n_1=1$, red squares)
{is automatically flattened out by the two-level multi-boson domain-decomposed HMC} ($n_1=10$, blue triangles)
without the need for modeling the long-distance behaviour of $G^{\rm conn}_{u,d}(x_0)$.

To further appreciate the effect of the two-level integration, we compute the integral in Eq.~(\ref{eq:amuint}) as
a function of the upper extrema of integration $x_0^{\rm max}$ which we allow to vary. For $n_1=1$, the
integral reads $446(26)$ and $424(38)$ for $x_0^{\rm max}=2.5$ and $3.0$~fm respectively, while
for $n_1=10$ the analogous values are $467.0(8.4)$ and $473.4(8.6)$. While with the one-level integration the errors
on the contributions to the integral from $0$ to $2.5$~fm and from $2.5$ to the maximum value of $3.0$~fm
are comparable, {with the two-level HMC the contribution to the variance from the long distance part becomes
negligible}. This pattern of variance reduction is expected to set in at shorter
distances for lighter quark masses, where the gain due to the two-level integration is
expected to be significantly larger due to the sharp increase of
the exponential in Eq.~(\ref{eq:relerr}). Considerations analogous to those made for the connected
contribution apply also to the much smaller disconnected one, although even larger values
of $n_1$ are required to render the variance approximately constant.

\section{Results and discussion}
Our best result for the light-connected contribution to the integrand in Eq.~(\ref{eq:amuint})
is shown on the left plot of Fig.~\ref{fig:integrand} (red squares). It is obtained
by a weighted average of the above discussed correlation function computed 
on $32$ point sources per time-slice on $7$ time-slices at $y_0/a=\{8,16,24,56,64,72,80\}$ and on $216$ sources at $y_0/a=32$.
We obtain a good statistical signal up to the maximum distance of $3$~fm or so.
The strange-connected contraction $G^{\rm conn}_{s}(x_0)$ is much less noisy, and is determined
by averaging on $16$ point sources at $y_0/a=32$. Its value, shown on the left plot
of Fig.~\ref{fig:integrand} (blue circles), is at most one order of magnitude smaller than the
light contribution, and has a negligible statistical error with respect to the light one.
The best result for the disconnected contribution has been computed as discussed in the previous section, and
it is shown in the left plot of Fig.~\ref{fig:integrand} as well (green triangles). It reaches a negative peak at
about $1.5$~fm, and a good statistical signal
is obtained up to $2.0$~fm or so. Its absolute value is more than two orders of magnitude smaller than the light-connected
contribution over the entire range explored (notice the multiplication by $10$ for a  better readability of the plot).\\
\indent In the right plot of Fig.~\ref{fig:integrand} we show the best values of the light-connected (red squares),
strange-connected (blue circles), and disconnected (green triangles) contributions to $a_\mu^{\rm HVP}\cdot 10^{10}$
as a function of the upper extrema of integration $x_0^{\rm max}$ in Eq.~(\ref{eq:amuint}).
The light-connected part starts to flatten out at $x_0^{\rm max} \sim 2.5$~fm and, at the conservative distance of
$x_0^{\rm max}= 3.0$~fm, its value is $471.8(6.2)$. The value of the strange-connected contribution is $52.55(21)$
at $x_0^{\rm max}=3.0$~fm, and its error is negligible with respect to the light-connected one.
The disconnected contribution starts to flatten out at about $x_0^{\rm max} \sim 2.0$~fm, where its value
is $-1.98(84)$. For $x_0^{\rm max} =3.0$~fm, its statistical uncertainty is $2.1$ which is still
3 times smaller with respect to the light-connected one. Clearly the disconnected contribution must be taken
into account to attain a per-mille precision on the HVP, but the combined usage of split-even
estimators and two-level integration solves the problem of its computation. By combining the connected
contributions at $x_0^{\rm max}=3.0$~fm with the disconnected part at $x_0^{\rm max}= 2.0$~fm, the best total
value that we obtain is $a_\mu^{\rm HVP} = 522.4(6.2) \cdot 10^{-10}$.\\[0.25cm]
\indent In this proof of concept study we have achieved a 1\% statistical precision with just
$n_0\cdot n_1=250$ configurations on a realistic lattice. This shows that for this light-quark mass
a per-mille statistical precision
on $a_\mu^{\rm HVP}$ is reachable with multi-level integration by increasing $n_0$ and $n_1$ by a factor of about
$4$--$6$ and $2$--$4$ respectively. When the up and the down quarks become lighter, the gain due
to the multi-level integration is expected to increase exponentially in the quark mass,
hence improving even more dramatically the scaling of the simulation cost with respect to a
standard one-level Monte Carlo. The change of computational paradigm presented here thus removes
the main barrier for making affordable, on the computer installations
available today, the goal of a per-mille precision on $a_\mu^{\rm HVP}$.\\
\indent Here we focused on the main bottleneck in the computation of the HVP.
It goes without saying that the very same variance-reduction pattern is expected to work out
also for the calibration of the
lattice spacing, the calculation of the electromagnetic
corrections and the HLbL.\\
\indent It is also interesting to notice that multi-level integration can be well integrated with master-field
simulation techniques~\cite{Francis:2019muy} if very large volumes turn out to be necessary to pin down
finite-size effects at the per-mille level. As a final remark, we stress that the very same approach is
applicable to many other computations which suffer from signal-to-noise ratio problems,
where a similar breakthrough is expected~\cite{Giusti:2017ksp}.\\

\section{Acknowledgments}
  \vspace{-0.25cm}
  
The generation of the configurations and the measurement of the correlators have been performed on the
PC clusters Marconi at CINECA (CINECA- INFN, CINECA-Bicocca agreements, ISCRA B project HP10BF2OQT)
and at the Juelich Supercomputing
Centre, Germany (PRACE project n. 2019215140) while the R\&D has been carried out on Wilson and Knuth at
Milano-Bicocca. We thank these institutions and PRACE for the computer resources and the
technical support. We also acknowledge PRACE for awarding us access to MareNostrum at Barcelona Supercomputing
Center (BSC), Spain (n. 2018194651) where comparative performance tests of the code have been performed.
We acknowledge partial support by the INFN project ``High performance data network''.

\bibliographystyle{elsarticle-num}
\bibliography{mb}
\end{document}